\documentclass[aps,pra,preprint]{revtex4-1}
\usepackage{hyperref}
\usepackage{amsmath}
\usepackage{amssymb}
\usepackage[pdftex]{graphicx}
\usepackage{gentium}
\usepackage{microtype}
\usepackage{verbatim}
\usepackage{parskip}

\makeatletter
\DeclareRobustCommand{\element}[1]{\@element#1\@nil}
\def\@element#1#2\@nil{%
  #1%
  \if\relax#2\relax\else\MakeLowercase{#2}\fi}
\makeatother

\newcommand{\mrm}{\mathrm}

\newcommand{\ket}[1]{{\left| {#1} \right\rangle}}
\newcommand{\bra}[1]{{\left\langle {#1} \right|}}

\newcommand{\braket}[2]{{\left\langle {#1}|{#2} \right\rangle}}
\newcommand{\avg}[1]{\left\langle {#1} \right\rangle }
\newcommand{\vecb}[1]{\mathbf{{#1}}}

\begin{document}
\title{Doppler and collisional frequency shifts in trapped-atom clocks}
\author{Amar C. Vutha}
\affiliation{Department of Physics, University of Toronto, Toronto, Canada ON M5S 1A7}
\email{vutha@physics.utoronto.ca}
\author{Tom Kirchner}
\affiliation{Department of Physics \& Astronomy, York University, Toronto, Canada M3J 1P3}
\author{Pierre Dub\'{e}}
\affiliation{National Research Council Canada, Ottawa, Canada K1A 0R6}
\date{\today}
\begin{abstract}

Collisions with background gas particles can shift the resonance frequencies of atoms in atomic clocks. The internal quantum states of atoms can also become entangled with their motional states due to the recoil imparted by a collision, which leads to a further shift of the clock frequency through the relativistic Doppler shift. It can be complicated to evaluate the Doppler and collisional frequency shifts for clock atoms in such entangled states, but estimates of these shifts are essential in order to improve the accuracy of optical atomic clocks. We present a formalism that describes collisions and relativistic Doppler shifts in a unified manner, and can therefore be used to accurately estimate collisional frequency shifts in trapped-atom clocks. 
% Here we develop a general collisional frequency shift formalism that properly accounts for the motion of trapped atoms in atomic clocks. Using this we evaluate the complete collisional frequency shift, including recoil effects, for a trapped-ion clock.
\end{abstract}
\maketitle

\section{Introduction}
Optical atomic clocks continue to make steady improvements in precision and accuracy \cite{Brewer2017,Campbell2017,Ludlow2018}. In the latest generation of trapped-atom (neutral and ion) clocks, the systematic uncertainty due to collisional shifts of the resonance frequency of clock atoms is a sizeable contribution to the uncertainty budget, typically at the level of $\sim$1 part in 10$^{18}$. In a previous paper \cite{Vutha2017}, we developed a method to estimate of the collisional frequency shift with improved accuracy -- for Sr$^+$ clocks for example, the frequency shift was calculated to be more than an order of magnitude smaller than earlier conservative estimates. Similar conclusions are likely to hold true for other trapped-ion clocks \cite{Davis2018}.

The analysis of collisional shifts in Ref.\ \cite{Vutha2017} relied on a simple framework that neglected (i.e., traced over) the motional degree of freedom of the clock atoms. Here we develop a formalism that describes the motion of a clock atom in a trap and the dynamics of its internal states in one unified framework. To keep the discussion simple, we model the clock atom as a system with two internal states $\ket{g},\ket{e}$, although the method can be easily extended to multi-level systems. The motivation to extend the formalism in Ref.\ \cite{Vutha2017} to include the motion of trapped atoms is as follows. The recoil momentum imparted to a clock atom during a collision depends on the scattering amplitude, which in turn depends on the interaction potential between the clock atom and a background gas particle, and therefore on the specific internal state of the atom. So a clock atom after a collision is typically left in an entangled state of its internal and motional degrees of freedom, of the form $\ket{\Psi} = c_g \ket{g} \ket{\psi_g} + c_e \ket{e} \ket{\psi_e}$, where $\ket{\psi_g},\ket{\psi_e}$ are motional states in the trap. Moving atoms experience a relativistic Doppler shift (also referred to as a second-order Doppler shift, or time-dilation shift) relative to the lab frame. In spite of the fact that atoms in optical clocks are cooled to sub-mK temperatures, the extreme accuracy of optical clocks means that even the small relativistic effects at these temperatures are nevertheless substantial compared to present-day clock performance. Doppler and collisional frequency shifts are therefore quite important as optical clock accuracies advance beyond 10$^{-18}$, and it is essential to be able to evaluate them accurately.

The evaluation of the relativistic Doppler shift usually proceeds by Lorentz transforming into the frame of the moving atom, followed by an expansion of the Lorentz $\gamma$ factor to leading order \cite{Fisk1997}. However, Lorentz transformation into the rest frame of a clock atom, which undergoes accelerated motion with zero mean velocity in a trap, is a questionable procedure at best. Furthermore, for entangled states such as $\ket{\Psi}$ above, it is not immediately evident which (combination) of the two motional state velocities should be used to evaluate the relativistic Doppler shift. An unambiguous method to evaluate collisional frequency shifts in such cases would be useful for improving the performance of optical atomic clocks; this is especially the case for clocks using single trapped-ions, where direct experimental measurements of background gas collisional shifts are difficult due to the long integration times necessary to measure such small shifts. Aside from its practical relevance for metrology though, the analysis of Doppler shifts in entangled states of clock atoms is also an interesting problem that confronts a fundamental notion: the inseparability of internal and motional degrees of freedom in situations where relativistic effects are important. 

In Sections \ref{sec:preliminaries} and \ref{sec:hamiltonian} we construct the Hamiltonian for a trapped clock atom, which properly accounts for the relativistic Doppler shift even when the clock atom is in an entangled state of its internal and motional degrees of freedom. In Section \ref{sec:dissipation} we develop a master equation for describing these degrees of freedom of the clock atom during collisions with background gas particles, and consider some limiting cases of this master equation to examine the consequences for optical atomic clocks.

\section{Preliminaries}\label{sec:preliminaries}
In the rotating-wave approximation, the interaction Hamiltonian for laser-atom interactions is 
\begin{equation}\label{eq:int_hamiltonian}
H_\mrm{int} = \frac{\Omega_0}{2} \left( S_+ \, e^{i \vecb{k} \cdot \vecb{x}} \, a_\vecb{k}  + S_- \, e^{-i \vecb{k} \cdot \vecb{x}} \, a_\vecb{k}^\dagger \right).
\end{equation}
(We use units where $\hbar$=1 throughout.) The internal state raising operator $S_+$ has the matrix element $\bra{e}S_+\ket{g} = 1$, $a_\vecb{k}$ is the annihilation operator for the laser mode with wavevector $\vecb{k}$, $\Omega_0$ is the vacuum Rabi frequency, and $\vecb{x}$ is the position of the atom's center of mass. Since $e^{i \vecb{k} \cdot \vecb{x}}$ increments the momentum of a motional state by $\vecb{k}$, we denote it as a momentum raising operator $M_\vecb{k}^\dagger$. Acting on a momentum eigenstate $\ket{\vecb{p}}$, this operator yields $M_\vecb{k}^\dagger\ket{\vecb{p}} = \ket{\vecb{p} + \vecb{k}}$. Equivalently, $M_\vecb{k}^\dagger$ is a phase space displacement operator that translates the momentum by $\vecb{k}$. The interaction Hamiltonian in Eq.\ (\ref{eq:int_hamiltonian}) can be rewritten as
\begin{equation}
H_\mrm{int} = \frac{\Omega_0}{2} \left( S_+ \, M_\vecb{k}^\dagger \, a_\vecb{k}  + S_- \, M_\vecb{k} \, a_\vecb{k}^\dagger \right).
\end{equation}
We assume that the laser mode is in a large coherent state $\ket{\alpha}$ ($|\alpha|^2 \gg 1$), and consider a transition from the ground to the excited state of the clock atom.

The matrix element of the interaction Hamiltonian, for a transition from a state $\ket{g} \ket{\psi_g}$ to $\ket{e} \ket{\psi_e}$, is then
\begin{equation}\label{eq:af_matrix_element}
\bra{\psi_e}\bra{e}H_\mrm{int}\ket{g}\ket{\psi_g} = \frac{\Omega_R}{2} e^{i(E_e - E_g)t} e^{-i \omega t} \, \bra{\psi_e} M_\vecb{k}^\dagger \ket{\psi_g}.
\end{equation}
Here $\Omega_R \approx \Omega_0 |\alpha|$ is the Rabi frequency, $E_g (E_e)$ is the energy eigenvalue of the joint internal-motional ground (excited) state, and $\omega$ is the angular frequency of the laser. There are two aspects to this equation: first, the motional matrix element $\bra{\psi_e} M_\vecb{k}^\dagger \ket{\psi_g}$ determines the overlap of the $\vecb{k}$-displaced initial motional state with the final motional state -- this is just the usual Lamb-Dicke factor \cite{Fisk1997}. Second, it is evident that the observed resonance frequency is simply $\omega_0 = E_e - E_g$. In order to properly account for relativistic shifts of the energy eigenvalues $E_g,E_e$, we use the following procedure.

We denote the rest mass of a reference energy level (e.g., the ground state of the atom, or the ionization threshold) as $m_0$. The mass of a stationary atom in an internal state $\ket{s}$ is $m_s = m_0 + \omega_s/c^2$, where $\omega_s$ is the energy of state $s$ relative to the reference energy level. Equivalently, $m_s c^2$ is the energy of the internal state $\ket{s}$ above the vacuum. The prescription for calculating the energy $E_s$ of an atom in a joint internal-motional state $\ket{s}\ket{\psi_s}$ is that it is \emph{just the energy of the motional state, but evaluated using the mass $m_s$} \cite{Dehn1970}\footnote{This approach to the relativistic Doppler shift is usually called the mass-change shift or mass defect, cf.\ \cite{Dehn1970,Keller2018,Yudin2018}, pioneered in the context of M{\"o}ssbauer spectroscopy in Ref.\ \cite{Josephson1960}.}. In the following, we illustrate this approach using the example of an atom in a harmonic trap (see Appendix \ref{appendix:free_atom_doppler} for the example of a free atom).

\subsection{Doppler shifts for a harmonically trapped atom}
The energy of the $n$-th eigenstate of a relativistic harmonic oscillator with mass $m$ and frequency $\sigma$ is (cf.\ \cite{Harvey1972}, Eq.\ (22))
\begin{equation}
E_n = mc^2 + \sigma \left[n+\frac{1}{2} + \frac{3}{4} \chi\left(n^2 + n + \frac{1}{2}\right) \right] + \mathcal{O}(\chi^2)
\end{equation}
where $\chi = \frac{\sigma}{4 mc^2}$. Even for trapped-ion clocks, where the typical trap secular frequencies ($\sigma \approx 2\pi \times$ 1 MHz) are higher than in neutral atom clocks, the quantity $\chi \sigma \sim 2\pi \times 10^{-12}$ Hz is a negligible frequency shift at the present level of clock accuracy. So it is adequate to approximate the energy of the $n$-th harmonic oscillator eigenstate as $E_n = mc^2 + \sigma \left(n+\frac{1}{2}\right)$. 

To explicitly show the mass (and therefore internal state) dependence of $\sigma$, we write $\sigma = \sqrt{\frac{\xi}{m}}$ where $\xi$ is the spring constant for the harmonic trap. If we assume a trap with a spring constant $\xi$ that is independent of the internal state of the atom, then the oscillation frequency in the trap for an internal state $\ket{s}$ is $\sigma_s = \sqrt{\frac{\xi}{m_s}} \approx \sigma_0 \left(1 - \frac{\omega_s}{2 m_0c^2} \right)$, with $\sigma_0 = \sqrt{\frac{\xi}{m_0}}$. 

As a result, the resonance frequency for a transition between $\ket{g}\ket{\psi_g}$ and $\ket{e}\ket{\psi_e}$ is 
%\begin{widetext}
\begin{equation}\label{eq:harmonic_trap}
\begin{split}
\omega_0 & = \left[mc^2 + \sigma \left(n+\frac{1}{2}\right)\right]_e - \left[mc^2 + \sigma \left(n+\frac{1}{2}\right)\right]_g \\
& \approx (\omega_e - \omega_g) + \sigma_0 \left(\avg{n}_e - \avg{n}_g\right) - \frac{\sigma_0}{2 m_0 c^2} \left[ \left(\avg{n}_e+\frac{1}{2}\right) \omega_e   - \left(\avg{n}_g+\frac{1}{2}\right) \omega_g  \right],
\end{split}
\end{equation}
%\end{widetext}
where $\avg{n}_s = \bra{\psi_s} \hat{n} \ket{\psi_s}$ is the expectation value of the number operator $\hat{n}$ of the oscillator. The first two terms are the internal and motional energy differences respectively, while the last term is the relativistic Doppler shift to leading order, which depends on a combination of the internal and motional energies. Expectedly, the relativistic Doppler shift persists in the limit when $\avg{n}_g = \avg{n}_e = 0$, due to the zero-point fluctuations in the harmonic oscillator. %The line strength of the transition between $\ket{g}\ket{\psi_g}$ and $\ket{e}\ket{\psi_e}$ is proportional to the squared Lamb-Dicke factor $|\bra{\psi_e}M_\vecb{k}^\dagger \ket{\psi_g}|^2$.

\emph{Radio-frequency ion trap.} The assumption of a fixed spring constant is not appropriate for rf ion traps, where the secular frequency of ion motion in the trap is not as simply dependent on mass as in an ideal harmonic oscillator. Typically $\sigma(m) = \sqrt{\frac{\alpha}{m} + \frac{\beta^2}{m^2}}$ for constants $\alpha,\beta$ that are determined by the ion trap's operating parameters \cite{Berkeland1998}. To express the relativistic Doppler shift for such cases in close analogy with the simple harmonic oscillator, we expand the trap frequency to leading order in the internal state energy and express it as
\begin{equation}
\sigma_s = \sigma(m_s) = \sigma_0 \left( 1 + \frac{\omega_s}{m_0c^2} \tau \right) + \mathcal{O}\left( \frac{\omega_s^2}{m_0^2c^4} \right),
\end{equation}
where we have defined the dimensionless quantity $\tau = \frac{m_0}{\sigma_0} \left( \frac{\partial \sigma}{\partial m} \right)_{m=m_0}$. For example $\tau = -\frac{1}{2}$ for the case of a fixed spring constant, and $\tau \approx -1$ for an rf ion trap with $a \ll q^2$ (where $a,q$ are the Mathieu parameters for the trap \cite{Berkeland1998}). This approach offers a simple way to extend calculations of the relativistic Doppler shift (such as those in, e.g., in Ref.\ \cite{Keller2018}) to ion traps where it cannot be assumed that $a \ll q^2$.

The measured resonance frequency can now be written as
%\begin{widetext}
\begin{equation}
\begin{split}
\omega_0 = (\omega_e - \omega_g) + \sigma_0 \left(\avg{n}_e - \avg{n}_g\right) + \tau \frac{\sigma_0}{m_0 c^2} \left[ \left(\avg{n}_e+\frac{1}{2}\right) \omega_e   - \left(\avg{n}_g+\frac{1}{2}\right) \omega_g  \right].
\end{split}
\end{equation}
%\end{widetext}

This expression is easy to generalize to a three dimensional ion trap with secular frequencies $\sigma_i(m_s) \approx \sigma_{0,i} + \frac{\omega_s}{m_0c^2} \tau_i$ ($i = x,y,z$) along three decoupled axes. The resonance frequency for a transition between $\ket{g}\ket{\psi_g}$ and $\ket{e}\ket{\psi_e}$ in such a case is
\begin{equation}\label{eq:ion_trap}
\begin{split}
\omega_0 = & (\omega_e - \omega_g) + \sum_i \sigma_{0,i} \left(\avg{n}_e - \avg{n}_g\right)_i \\ 
& + \sum_i \tau_i \frac{\sigma_{0,i}}{m_0 c^2} \left[ \left(\avg{n}_e+\frac{1}{2}\right) \omega_e   - \left(\avg{n}_g+\frac{1}{2}\right) \omega_g  \right]_i.
\end{split}
\end{equation}
The last term is the relativistic Doppler shift in an rf ion trap. 

\section{Hamiltonian for a trapped-atom clock}\label{sec:hamiltonian}
Based on the preceding sections, we can write the Hamiltonian operator for a harmonically trapped two-level atom interacting with a laser as
\begin{equation}\begin{split}
H = & \ket{g}\bra{g} \left[ m_g c^2 + \sigma_g \left(\hat{n} + 1/2\right) \right] + \ket{e}\bra{e} \left[ m_e c^2 + \sigma_e \left(\hat{n} + 1/2\right) \right] \\ 
& + \frac{\Omega_R}{2} \left( \ket{g}\bra{e} M_\vecb{k} e^{i\omega t} + \ket{e}\bra{g} M_\vecb{k}^\dagger e^{-i\omega t} \right).
\end{split}\end{equation}

For ease of computation in the following sections, we denote this Hamiltonian as a matrix in the $\ket{g}, \ket{e}$ basis, with entries that are \emph{operators} acting on the state space of the motional degrees of freedom.
%\begin{widetext}
\begin{equation}\begin{split}
H & = \begin{pmatrix}
m_g c^2 \mathbb{I} + \sigma_g \left(\hat{n} + \mathbb{I}/2\right) & \frac{\Omega_R}{2} e^{i \omega t} M_\vecb{k} \\
\frac{\Omega_R}{2} e^{-i \omega t} M_\vecb{k}^\dagger & m_e c^2 \mathbb{I} + \sigma_e \left(\hat{n} + \mathbb{I}/2\right) 
\end{pmatrix} \\
& = \begin{pmatrix} (m_g c^2  + \frac{1}{2} \sigma_g)\mathbb{I} & 0 \\ 0 & (m_e c^2 + \frac{1}{2}\sigma_e)\mathbb{I} \end{pmatrix} + \begin{pmatrix} \sigma_g  \hat{n}& 0 \\ 0 & \sigma_e \hat{n} \end{pmatrix} + \begin{pmatrix} 0 & \frac{\Omega_R}{2} e^{i\omega t} M_\vecb{k} \\ \frac{\Omega_R}{2} e^{-i\omega t} M_\vecb{k}^\dagger & 0 \end{pmatrix}.
\end{split}\end{equation}
%\end{widetext}

Defining the ideal resonance frequency $\omega_{00} = (m_e c^2  + \frac{1}{2} \sigma_e) - (m_g c^2  + \frac{1}{2} \sigma_g)$, and the detuning $\Delta = \omega - \omega_{00}$, the usual transformation to a basis of states dressed by the laser %\cite{Cohen-Tannoudji} 
(equivalently here, the semiclassical rotating frame transformation $\ket{g} \to e^{i \omega t/2} \ket{g}, \ket{e} \to e^{-i \omega t/2}\ket{e}$) allows us to reduce the Hamiltonian to the simple form
\begin{equation}
H = \begin{pmatrix} \frac{\Delta}{2} & 0 \\ 0 & -\frac{\Delta}{2} \end{pmatrix} + \begin{pmatrix} \sigma_g  \hat{n}& 0 \\ 0 & \sigma_e \hat{n} \end{pmatrix} + \begin{pmatrix} 0 & \frac{\Omega_R}{2} M_\vecb{k} \\ \frac{\Omega_R}{2} M_\vecb{k}^\dagger & 0 \end{pmatrix}.
\end{equation}

We write the joint internal-motional density operator $\rho$ of the trapped atom as a matrix in the same $\ket{g},\ket{e}$ basis, $\rho = \begin{pmatrix} \bar{\rho}_{gg} & \bar{\rho}_{ge} \\ \bar{\rho}_{eg} & \bar{\rho}_{ee} \end{pmatrix}$,
where the matrix elements are again understood to be operators in the motional state space. (For example, $\bar{\rho}_{ge} = \bra{g} \rho \ket{e}$, and $\bar{\rho}_{eg} = \bar{\rho}_{ge}^\dagger$.) The unitary evolution of the density matrix is determined by $\frac{d\rho}{dt} = -i[H,\rho]$, which in matrix form is
%\begin{widetext}
\begin{equation}
\begin{split}\label{eq:schrodinger}
\frac{d}{dt}\begin{pmatrix} \bar{\rho}_{gg} & \bar{\rho}_{ge} \\ \bar{\rho}_{eg} & \bar{\rho}_{ee} \end{pmatrix}_\mrm{unitary} = & -i \begin{pmatrix} 0 & \Delta \bar{\rho}_{ge} \\ -\Delta \bar{\rho}_{eg} & 0 \end{pmatrix} -i  \begin{pmatrix} \sigma_g [\hat{n},\bar{\rho}_{gg}] & \sigma_g \hat{n} \bar{\rho}_{ge} - \sigma_e \bar{\rho}_{ge} \hat{n} \\ \sigma_e \hat{n} \bar{\rho}_{eg} - \sigma_g \bar{\rho}_{eg} \hat{n} & \sigma_e [\hat{n},\bar{\rho}_{ee}] \end{pmatrix} \\
& -i \frac{\Omega_R}{2}\begin{pmatrix} M_\vecb{k} \bar{\rho}_{eg} - \bar{\rho}_{ge}M_\vecb{k}^\dagger & M_\vecb{k} \bar{\rho}_{ee} - \bar{\rho}_{gg} M_\vecb{k} \\ M_\vecb{k}^\dagger \bar{\rho}_{gg} - \bar{\rho}_{ee} M_\vecb{k}^\dagger & M_\vecb{k}^\dagger \bar{\rho}_{ge} - \bar{\rho}_{eg}M_\vecb{k} \end{pmatrix}.
\end{split}
\end{equation}
%\end{widetext}

\section{Effect of collisions}\label{sec:dissipation}
We model elastic collisions between the trapped clock atom and a background gas particle using a set of Lindblad jump operators acting on the joint internal-motional states of the clock atom,
\begin{equation}\label{eq:lindblad_operators}
L_{\vecb{p}',\vecb{p}} = \sqrt{\frac{n p}{\mu}} \begin{pmatrix} f_g(\vecb{p}',\vecb{p}) \, M_\vecb{q}^\dagger & 0 \\ 0 & f_e(\vecb{p}',\vecb{p}) \, M_\vecb{q}^\dagger \end{pmatrix},
\end{equation}
where $n$ is the number density of the background gas, $\mu$ is the reduced mass of the colliding particles, $f_s(\vecb{p}',\vecb{p})$ is the scattering amplitude for a collision with momentum $\vecb{p} \to \vecb{p'}$ (in the center of mass frame of the colliding particles) when the clock atom is in an internal state $\ket{s}$, and $p = |\vecb{p}| = |\vecb{p}'|$. The unitary operator $M_\vecb{q}^\dagger$ acts on the motional degree of freedom of the clock atom and displaces it by the recoil momentum, $\vecb{q} = \vecb{p} - \vecb{p}'$. We have assumed that collisions do not change the internal states of the clock atom. The form of these jump operators is derived in Appendix \ref{appendix:jump_operator}.

For a background gas collision with momentum $\vecb{p}$, the dissipative part of the density matrix equation of motion is obtained by summing the effect of jump operators over all possible directions of the momentum $\vecb{p}'$.
%\begin{widetext}
\begin{equation}
\left(\frac{d \rho}{d t}\right)_\mrm{diss} = \int d\Omega(\vecb{p}') \, \left[ L_{\vecb{p}',\vecb{p}} \, \rho \, L_{\vecb{p}',\vecb{p}}^\dagger - \frac{1}{2} \left( L_{\vecb{p}',\vecb{p}}^\dagger L_{\vecb{p}',\vecb{p}}\, \rho + \rho \, L_{\vecb{p}',\vecb{p}}^\dagger L_{\vecb{p}',\vecb{p}} \right) \right].
\end{equation}
%\end{widetext}
For collisions with a background gas ensemble, the right hand side of the above equation can be averaged over the distribution of the collision momenta, $N(\vecb{p})$, to obtain
%\begin{widetext}
\begin{equation}
\left(\frac{d \rho}{d t}\right)_\mrm{diss} = \int d^3\vecb{p} \, N(\vecb{p}) \int d\Omega(\vecb{p}') \, \left[ L_{\vecb{p}',\vecb{p}} \, \rho \, L_{\vecb{p}',\vecb{p}}^\dagger - \frac{1}{2} \left( L_{\vecb{p}',\vecb{p}}^\dagger L_{\vecb{p}',\vecb{p}}\, \rho + \rho \, L_{\vecb{p}',\vecb{p}}^\dagger L_{\vecb{p}',\vecb{p}} \right) \right].
\end{equation}
%\end{widetext}

Inserting the matrix forms of $L_{\vecb{p}',\vecb{p}}$ and $\rho$ leads to the equation
%\begin{widetext}
\begin{equation}\label{eq:lindblad}
\begin{split}
\frac{d}{dt}\begin{pmatrix} \bar{\rho}_{gg} & \bar{\rho}_{ge} \\ \bar{\rho}_{eg} & \bar{\rho}_{ee} \end{pmatrix}_\mrm{diss} = \int d^3\vecb{p}\,  N(\vecb{p}) \frac{n p}{\mu} \int d\Omega(\vecb{p}')  & \left[ \begin{pmatrix} |f_g|^2 M_\vecb{q}^\dagger \, \bar{\rho}_{gg} M_\vecb{q} & f_g f_e^* M_\vecb{q}^\dagger \, \bar{\rho}_{ge} M_\vecb{q} \\ f_e f_g^* M_\vecb{q}^\dagger \, \bar{\rho}_{eg} M_\vecb{q} & |f_e^2| M_\vecb{q}^\dagger \, \bar{\rho}_{ee} M_\vecb{q} \end{pmatrix} \right.\\
& - \left. \begin{pmatrix} |f_g|^2 \bar{\rho}_{gg} & \left(\frac{|f_g|^2 + |f_e|^2}{2}\right) \bar{\rho}_{ge} \\ \left(\frac{|f_g|^2 + |f_e|^2}{2}\right) \bar{\rho}_{eg} & |f_e|^2 \bar{\rho}_{ee} \end{pmatrix} \right].
\end{split}
\end{equation}
%\end{widetext}
We have used the symbols $f_g, f_e$ as shorthand for $f_g(\vecb{p}',\vecb{p}), f_e(\vecb{p}',\vecb{p})$ respectively. 

The full equation of motion of the joint internal-motional density matrix is the sum of Eqs.\ (\ref{eq:schrodinger}) and (\ref{eq:lindblad}):
%\begin{widetext}
\begin{equation}\label{eq:full_eom}
\begin{split}
\frac{d}{dt}\begin{pmatrix} \bar{\rho}_{gg} & \bar{\rho}_{ge} \\ \bar{\rho}_{eg} & \bar{\rho}_{ee} \end{pmatrix} = & -i \begin{pmatrix} 0 & \Delta \bar{\rho}_{ge} \\ -\Delta \bar{\rho}_{eg} & 0 \end{pmatrix} -i  \begin{pmatrix} \sigma_g [\hat{n},\bar{\rho}_{gg}] & \sigma_g \hat{n} \bar{\rho}_{ge} - \sigma_e \bar{\rho}_{ge} \hat{n} \\ \sigma_e \hat{n} \bar{\rho}_{eg} - \sigma_g \bar{\rho}_{eg} \hat{n} & \sigma_e [\hat{n},\bar{\rho}_{ee}] \end{pmatrix} \\
& -i \frac{\Omega_R}{2}\begin{pmatrix} M_\vecb{k} \bar{\rho}_{eg} - \bar{\rho}_{ge}M_\vecb{k}^\dagger & M_\vecb{k} \bar{\rho}_{ee} - \bar{\rho}_{gg} M_\vecb{k} \\ M_\vecb{k}^\dagger \bar{\rho}_{gg} - \bar{\rho}_{ee} M_\vecb{k}^\dagger & M_\vecb{k}^\dagger \bar{\rho}_{ge} - \bar{\rho}_{eg}M_\vecb{k} \end{pmatrix} \\
& + \int d^3\vecb{p}\,  N(\vecb{p}) \frac{n p}{\mu} \int d\Omega(\vecb{p}')  \left[ \begin{pmatrix} |f_g|^2 M_\vecb{q}^\dagger \, \bar{\rho}_{gg} M_\vecb{q} & f_g f_e^* M_\vecb{q}^\dagger \, \bar{\rho}_{ge} M_\vecb{q} \\ f_e f_g^* M_\vecb{q}^\dagger \, \bar{\rho}_{eg} M_\vecb{q} & |f_e^2| M_\vecb{q}^\dagger \, \bar{\rho}_{ee} M_\vecb{q} \end{pmatrix} \right.\\
& \ \ \ \ \ \ \ \ \ \ \ \ \ \ \ \ \ \ \ \ \ \ \ \ \ \ \ \ \ \ \ \ \ \ \ \ \ \ \ \ \ \ \ \ \ \ \ \ \ \ \ -\left. \begin{pmatrix} |f_g|^2 \bar{\rho}_{gg} & \left(\frac{|f_g|^2 + |f_e|^2}{2}\right) \bar{\rho}_{ge} \\ \left(\frac{|f_g|^2 + |f_e|^2}{2}\right) \bar{\rho}_{eg} & |f_e|^2 \bar{\rho}_{ee} \end{pmatrix} \right].
\end{split}
\end{equation}
%\end{widetext}
The terms on the right hand side are respectively the detuning term, the relativistic Doppler shift due to motion in the trap, the driving term due to the laser, and the dissipative term due to collisions. This equation is the central result of this paper.

When precise estimates of collision-induced shifts are required, Eq.\ (\ref{eq:full_eom}) for the density matrix has to be solved numerically in general. (The necessary scattering amplitudes $f_{g,e}(\vecb{p}',\vecb{p})$ can be obtained from \emph{ab initio} calculations of molecular potential energy curves \cite{Vutha2017}.) In addition to collisional frequency shifts, the solution to this equation will also describe processes such as collisional recoil of the clock atom out of the initial motional state into other motional states, and the resulting modifications of the relativistic Doppler shift of the resonance. Numerical calculations of these effects will be explored in a forthcoming publication.

Nevertheless, although they do not admit general analytic solutions, some instructive conclusions can still be drawn from Eq.\ (\ref{eq:lindblad}) and Eq.\ (\ref{eq:full_eom}) in a couple of limiting cases:
\begin{enumerate}
\item \emph{Trace over motional states:} When the only pieces of information available in a measurement are the clock atom internal state population and coherences, the density matrix is traced over the motional degrees of freedom of the atom. Setting aside the driving term due to the laser, the trace of Eq.\ (\ref{eq:full_eom}) over motional degrees of freedom yields 
%\begin{widetext}
\begin{equation}\label{eq:lindblad_trace}
\begin{split}
\frac{d}{dt}\begin{pmatrix} {\rho}_{gg} & {\rho}_{ge} \\ {\rho}_{eg} & {\rho}_{ee} \end{pmatrix} = - & i \begin{pmatrix} 0 & \Delta {\rho}_{ge} + (\sigma_g - \sigma_e) \mrm{tr}(\hat{n} \bar{\rho}_{ge}) \\ -\Delta {\rho}_{eg} + (\sigma_e - \sigma_g) \mrm{tr}(\hat{n} \bar{\rho}_{eg}) & 0 \end{pmatrix} \\
& + \begin{pmatrix} 0 & \kappa {\rho}_{ge} \\ \kappa^* {\rho}_{eg}  & 0 \end{pmatrix},
\end{split}
\end{equation}
%\end{widetext}
where, e.g., $\rho_{gg}, \rho_{eg}$ etc. are now scalars representing the density matrix elements in the internal state space. The total population in the $g,e$ internal states is unchanged by the collision, consistent with our assumption of elastic collisions. The rate at which the coherence $\rho_{ge}$ evolves due to collisions is
\begin{equation}
\kappa = \int d^3\vecb{p}\,  N(\vecb{p}) \frac{n p}{\mu} \int d\Omega(\vecb{p}')  \left[ f_g f_e^* - \left(\frac{|f_g|^2 + |f_e|^2}{2}\right) \right].
\end{equation}
In Eq.\ (\ref{eq:lindblad_trace}) $i\kappa$ appears on equal footing with the detuning $\Delta$, allowing us to read off the collisional frequency shift (CFS). This leads to the expression for the CFS correction (collision-free resonance frequency minus measured resonance frequency) that was derived in Ref.\ \cite{Vutha2017} using a different approach:
%\begin{widetext}
\begin{equation}\label{eq:trace_cfs}
\delta \omega_\mrm{CFS} = - \mathfrak{Im} \int d^3\vecb{p}\,  N(\vecb{p}) \frac{n p}{\mu} \int d\Omega(\vecb{p}') \, f_g(\vecb{p}',\vecb{p}) f_e^*(\vecb{p}',\vecb{p}).
\end{equation}
%\end{widetext}

\item \emph{Projection onto motional ground state:} Another interesting limit is when the joint internal-motional state can be assumed to be $\ket{\Psi(0)} = ( c_g \ket{g} + c_e \ket{e} ) \ket{0}$, where $\ket{0}$ is the motional ground state in the trap. Taking the expectation value of Eq.\ (\ref{eq:lindblad}) in $\ket{0}$ models the dissipative dynamics of a clock with an atom cooled to its motional ground state (e.g., a trapped-ion clock with sideband cooling \cite{Brewer2017}), where the atomic populations and coherences in $\ket{0}$ can be selectively measured.

We consider the population in $\ket{g}\ket{0}$, which is $\rho_{gg,0} = \bra{0} \bar{\rho}_{gg} \ket{0} = |c_g|^2$. The typical collision rate in the ultrahigh-vacuum environment of a trapped-atom clock is $\sim 10^{-3}$/s \cite{Dube2013}. Since the typical measurement cycle in these clocks ($\sim$ 0.1-1 s) is short compared to the inverse collision rate, it is sufficient to use first-order perturbation theory for the dissipative part of the time evolution of $\rho_{gg,0}$. We therefore obtain
%\begin{widetext}
\begin{equation}
\left(\frac{d \rho_{gg,0}}{dt}\right)_\mrm{diss} \approx \int d^3\vecb{p}\,  N(\vecb{p}) \frac{n p}{\mu} \int d\Omega(\vecb{p}') \,  |f_g(\vecb{p}',\vecb{p})|^2 \, \left( |\bra{0} M_\vecb{q}^\dagger \ket{0}|^2  - 1  \right)  \, \rho_{gg,0}.
\end{equation}
%\end{widetext}
%where the time dependence of $\rho_{gg,0}$ on the right hand side can be approximated as arising from just the unitary dynamics.

The equation for the coherence $\rho_{ge,0} = \bra{0} \bar{\rho}_{ge} \ket{0} = c_g c_e^*$ under the same conditions is
%\begin{widetext}
\begin{equation}\label{eq:coherence_decay}
\left(\frac{d \rho_{ge,0}}{dt}\right)_\mrm{diss} \approx \int d^3\vecb{p}\,  N(\vecb{p}) \frac{n p}{\mu} \int d\Omega(\vecb{p}') \, \left[ f_g f_e^* \, |\bra{0} M_\vecb{q}^\dagger \ket{0}|^2  - \left(\frac{|f_g|^2 + |f_e|^2}{2} \right)  \right] \rho_{ge,0}.
\end{equation}
%\end{widetext}

%\frac{d \rho_{gg,0}}{dt} \approx \left[ \int d^3\vecb{p}\,  N(\vecb{p})\frac{np}{\mu} \int d\Omega \, |f_g(\vecb{p},\theta)|^2 \, \left( |\bra{0} M_{\vecb{p}(1-\cos \theta)}^\dagger \ket{0}|^2  - 1  \right)  \right] \rho_{gg,0}.
If the recoil momentum is large compared to the momentum uncertainty in the motional ground state, then the displaced state $M_\vecb{q}^\dagger \ket{0}$ has negligible overlap with $\ket{0}$, and $|\bra{0} M_\vecb{q}^\dagger \ket{0}|^2 \approx 0$. In this limit each collision effectively removes population out of the motional ground state, and the population $\rho_{gg,0}$ decays exponentially at the collision rate 
\begin{equation}
\gamma_g = \int d^3\vecb{p}\,  N(\vecb{p}) \frac{n p}{\mu} \int d\Omega(\vecb{p}') \, |f_g(\vecb{p}',\vecb{p})|^2.
\end{equation}
%\gamma_g = \int d^3\vecb{p}\,  N(\vecb{p}) \frac{np}{\mu} \int d\Omega \, |f_g(\vecb{p},\theta)|^2.
A similar expression involving the excited state collision rate $\gamma_e$ holds for $\rho_{ee,0}$, the population in $\ket{e}\ket{0}$, in this limit; furthermore, from Eq.\ (\ref{eq:coherence_decay}) the coherence $\rho_{ge,0}$ decays at the average of $\gamma_g$ and $\gamma_e$, and there is no imaginary term that would lead to a collisional frequency shift.

In a 3D harmonic trap with secular frequencies $\sigma_{0,i}$ ($i = x,y,z$), the overlap factor between the displaced state $M_\vecb{q}^\dagger\ket{0}$ and the ground state $\ket{0}$ (a collisional analog of the Lamb-Dicke factor) is
\begin{equation}
|\bra{0} M_\vecb{q}^\dagger \ket{0}|^2 \approx \exp\left[-\frac{1}{2} \sum_i \frac{(p_i - p'_i)^2}{m_0 \, \sigma_{0,i}}  \right].
\end{equation}
For atoms prepared and detected in $\ket{0}$, the CFS correction is 
%\begin{widetext}
\begin{equation}
\delta \omega_\mrm{CFS} \approx - \mathfrak{Im} \int d^3\vecb{p}\,  N(\vecb{p}) \frac{n p}{\mu} \int d\Omega(\vecb{p}') \, f_g(\vecb{p}',\vecb{p}) f_e^*(\vecb{p}',\vecb{p}) \, |\bra{0} M_\vecb{q}^\dagger \ket{0}|^2.
\end{equation}
%\end{widetext}
This correction can be considerably smaller than the result in Eq.\ (\ref{eq:trace_cfs}) when the overlap factor is small. Motional-state-resolved detection of clock atoms can therefore lead to significant suppression of collisional shifts.

\end{enumerate}

\section{Summary}
We have developed a unified formalism for describing relativistic Doppler shifts and collisional frequency shifts, which offers a systematic method to calculate collision-induced frequency shifts in trapped-atom clocks. This framework enables the accurate evaluation of systematic errors in clocks due to both the phase shifts, as well as the momentum recoils, imparted by background gas collisions. Our analysis shows that motional-state-resolved detection of clock transitions can be used to significantly suppress the collisional shift, indicating that the systematic uncertainty due to collisions can be reduced to negligible levels in clocks that use atoms cooled to their motional ground states. 

We acknowledge support from the Branco Weiss Fellowship, NSERC, and Canada Research Chairs.

\bibliography{crs.bib}

\appendix
\section{Relativistic Doppler shift for a free atom}\label{appendix:free_atom_doppler}
As an application of the methods used in this paper, we show how the relativistic Doppler shift reduces the familiar expression $\frac{\delta \omega_0}{\omega_0} = -\frac{1}{2}\frac{v^2}{c^2}$ for a freely moving atom. The energy of an atom in a momentum eigenstate $\ket{\vecb{p}_s}$, when it is in an internal state $s$, is $E_s = \sqrt{m_s^2 c^4 + p_s^2 c^2}$. In a transition between states $\ket{g}\ket{\vecb{p}_g}$ and $\ket{e}\ket{\vecb{p}_e}$, the motional part of the transition matrix element, $\bra{\psi_e} M_\vecb{k}^\dagger \ket{\psi_g} = \bra{\vecb{p}_e}M_\vecb{k}^\dagger \ket{\vecb{p}_g}$, is only nonzero when $\vecb{p}_e = \vecb{k} + \vecb{p}_g$, enforcing strict momentum conservation. The resonance frequency of the transition is therefore
\begin{equation}\label{eq:free_particle}
\begin{split}
\omega_0 & = \left[\sqrt{m^2 c^4 + p^2 c^2} \right]_e - \left[\sqrt{m^2 c^4 + p^2 c^2} \right]_g \\
& \approx (\omega_e - \omega_g) + \frac{\vecb{p}_g \cdot \vecb{k}}{m_e} + \frac{k^2}{2 m_e} -\frac{1}{2} \left(\frac{p_g}{m_0 c}\right)^2 (\omega_e - \omega_g),
\end{split}
\end{equation}
up to $\mathcal{O}\left[ \left(\frac{p_{g,e}}{m_0c}\right)^4,\left(\frac{\omega_{g,e}}{m_0c^2}\right)^2 \right]$. The corrections to the naive resonance frequency $(\omega_e-\omega_g)$ are respectively the usual expressions for the first-order Doppler, photon recoil and second-order Doppler shifts for an atom in free space.

\section{Jump operators for scattering}\label{appendix:jump_operator}
Consider a scattering process between a clock atom in a joint motional-internal state $\ket{\vecb{p}_a}\ket{\psi_\mrm{int}}$ and a background gas particle in a motional state $\ket{\vecb{p}_b}$. The joint state of the atom and background particle is $\ket{\vecb{p}_a}\ket{\vecb{p}_b}\ket{\psi_\mrm{int}}$, which can also be described in terms of the center of mass (CM) and relative motion degrees of freedom as $\ket{\vecb{P}_\mrm{CM}}\ket{\vecb{p}}\ket{\psi_\mrm{int}}$. Here the total momentum of the system is $\vecb{P}_\mrm{CM} = \vecb{p}_a + \vecb{p}_b$ and the collision momentum is $\vecb{p} = \mu\left(\frac{\vecb{p}_b}{m_b} - \frac{\vecb{p}_a}{m_a}\right)$. 

The eigenstate of the scattering Hamiltonian, in the asymptotic limit when the colliding particles are well separated, can be written as
%\begin{widetext}
\begin{equation}
\begin{split}
\ket{\Psi} & = \mathcal{A} \ket{\vecb{P}_\mrm{CM}}\left( \ket{\vecb{p}}\ket{\psi_\mrm{int}} + \int d\Omega(\vecb{p'}) \, \mathbb{F}(\vecb{p'},\vecb{p}) \ket{\vecb{p'}}\ket{\psi_\mrm{int}} \right) \\
& = \mathcal{A} \ket{\vecb{P}_\mrm{CM}}\left[ \mathbb{I} + \int d\Omega(\vecb{p'}) \, \left(M^{(r)\dagger}_{\vecb{p'}-\vecb{p}} \otimes \mathbb{F}(\vecb{p'},\vecb{p}) \right)  \right] \ket{\vecb{p}}\ket{\psi_\mrm{int}},
\end{split}
\end{equation}
%\end{widetext}
where $\mathcal{A}$ is a normalization factor and $\mathbb{F}(\vecb{p'},\vecb{p})$ is the $\vecb{p'} \leftarrow \vecb{p}$ scattering amplitude matrix that acts on the internal states. The matrix $\mathbb{F}(\vecb{p'},\vecb{p})$ is diagonal for elastic collisions, with diagonal elements $f_s(\vecb{p'},\vecb{p})$ ($s = g,e$). The scattering amplitude $f_s$ depends on the potential energy curve for the molecule formed between the clock atom in state $\ket{s}$ and the background gas particle. The momentum displacement operator $M^{(r)\dagger}_{\vecb{p'}-\vecb{p}}$ acts on the relative motion degree of freedom. The momentum of the center of mass is decoupled from the scattering process, and $\ket{\vecb{P}_\mrm{CM}}$ remains unaffected by the collision. Projecting $\ket{\Psi}$ on to an eigenket of the relative coordinate, $\ket{r\theta \phi}$, yields the usual form of the scattering wavefunction, $\braket{r \theta \phi}{\Psi} = \mathcal{A} \ket{\vecb{P}_\mrm{CM}}\left( e^{ipz} + \mathbb{F}(\theta,\phi) \frac{e^{ipr}}{r} \right) \ket{\psi_\mrm{int}}$.

In terms of lab frame momenta, $\ket{\Psi}$ is
%\begin{widetext}
\begin{equation}
\ket{\Psi} = \mathcal{A} \left[ \mathbb{I} + \int d\Omega(\vecb{p'}) \, \left(M^{(b)\dagger}_{\vecb{p'}-\vecb{p}} \otimes M^{(a) \dagger}_{\vecb{p} - \vecb{p'}}  \otimes \mathbb{F}(\vecb{p'},\vecb{p}) \right)  \right] \ket{\vecb{p}_b}\ket{\vecb{p}_a}\ket{\psi_\mrm{int}},
\end{equation}
%\end{widetext}
where $M^{(a) \dagger}_{\vecb{p} - \vecb{p'}}$ is the momentum recoil acting on the clock atom, and $M^{(b) \dagger}_{ \vecb{p' - \vecb{p}}}$ acts on the background gas particle. In this step, we have used the fact that differences between momenta are the same in the lab frame and CM frame. It is convenient to retain the integral over the outgoing momentum $\vecb{p'}$ for the relative motion of the two particles.

Following the procedure described in Ref.\ \cite{Vutha2017}, (i) we choose $\mathcal{A} = \sqrt{\frac{np \, \delta t}{\mu}}$, where $\delta t$ is assumed to be long compared to the duration of the collision but short compared to the mean time between collisions; and (ii) we construct a family of Kraus operators $K_{\vecb{p}_b',\vecb{p}}$ by projecting $\ket{\Psi}$ onto $\ket{\vecb{p}_b'}$, a complete basis for the motional states of the background gas particle. The Kraus operators project the unitary collision dynamics onto the subspace of interest (the clock atom's internal and motional states) \cite{Nielsen2011}. 

Each background particle momentum $\vecb{p}_b'$ is in one-to-one correspondence with the collision momentum $\vecb{p'} = \vecb{p} + \vecb{p}_b' - \vecb{p}_b$, and therefore the Kraus operators can equivalently be indexed using $\vecb{p'}$. The resulting Kraus operators, written explicitly as a matrix in the internal state space, are
\begin{equation}
K_{\vecb{p'},\vecb{p}} = \sqrt{\frac{n p \, \delta t}{\mu}} 
\begin{pmatrix} f_g(\vecb{p'},\vecb{p}) \, M_\vecb{p - p'}^\dagger &  0 \\ 0 & f_e(\vecb{p'},\vecb{p}) \, M_\vecb{p - p'}^\dagger \end{pmatrix}.
\end{equation}

We make the Markov assumption that the bath of colliding particles has zero correlations between consecutive collisions. This assumption is justified for a clock atom trapped inside a vacuum chamber, where the momenta of the background gas particles are randomized after each collision with the wall of the chamber. Therefore we combine the effect of the Kraus operators on the density matrix, rather than the wavefunction. To obtain a differential equation for the density matrix, we associate a Lindblad jump operator $L_\vecb{p',p} = K_\vecb{p',p}/\sqrt{\delta t}$ with each of the Kraus operators \cite{Vutha2017}. With the definition $\vecb{q} = \vecb{p} - \vecb{p'}$, we are led to the jump operators defined in Eq.\ (\ref{eq:lindblad_operators}):
\begin{equation}
L_{\vecb{p'},\vecb{p}} = \sqrt{\frac{n p}{\mu}} \begin{pmatrix} f_g(\vecb{p}',\vecb{p}) \, M_\vecb{q}^\dagger & 0 \\ 0  & f_e(\vecb{p}',\vecb{p}) \, M_\vecb{q}^\dagger \end{pmatrix}.
\end{equation}

For completeness, we note that there are situations where %the background gas bath cannot be assumed to be Markovian ??, and 
consecutive collisions are correlated (e.g., collisions of a moving clock atom with a background of scatterers, as in fountain clocks \cite{Gibble2013}). In such cases the action of the Kraus operators on the \emph{wavefunction}, instead of the density matrix, must be combined together. Additional frequency shifts can result in these cases -- for example, adding the wavefunction for multiple forward scattering events leads to the well-known matter wave refractive index effect \cite{Vigue1995,Champenois2008}.

\end{document}